\documentclass{article}

\usepackage{amsmath}
\usepackage{hyperref}
\usepackage{lineno}
\usepackage{multirow}
\usepackage{xcolor}
\usepackage{graphicx}
\usepackage{authblk}
\usepackage[a4paper, total={7in, 10in}]{geometry}

\title{A neuromorphic camera for tracking passive and active matter with lower data throughput}
\date{}

\author[1,2,3]{Gabriel Britto Monteiro}
\author[2,4]{Megan Lim}
\author[2,4]{Tiffany Cheow Yuen Tan}
\author[1,2]{Avinash Upadhya}
\author[5]{Zhuo Liang}
\author[5]{Benjamin Agnew}
\author[5]{Tomonori Hu}
\author[3,5]{Benjamin J. Eggleton}
\author[1,2,3]{Christopher Perrella}
\author[2,4]{Kylie Dunning}
\author[1,2,3,6,*]{Kishan Dholakia}

\affil[1]{Centre of Light for Life and School of Biological Sciences, The University of Adelaide, Australia}
\affil[2]{Institute for Photonics and Advanced Sensing, The University of Adelaide, Australia}
\affil[3]{ARC Centre of Excellence in Optical Microcombs for Breakthrough Science (COMBS)}
\affil[4]{Robinson Research Institute and School of Biomedicine, The University of Adelaide, Australia}
\affil[5]{Institute of Photonics and Optical Science (IPOS), School of Physics, The University of Sydney, Australia}
\affil[6]{SUPA, School of Physics and Astronomy, University of St Andrews, KY16 9SS, United Kingdom}
\affil[*]{Corresponding Author: kishan.dholakia@adelaide.edu.au}

\begin{document}

\maketitle

\begin{abstract}
   We demonstrate the merits of using a neuromorphic, or event-based camera (EBC), for tracking of both passive and active matter. For passive matter, we tracked the Brownian motion of different micro-particles and estimated their diffusion coefficient. For active matter, we explored the case of tracking murine spermatozoa and extracted motility parameters from the motion of cells. This has applications in enhancing outcomes for clinical fertility treatments. Using the EBC, we obtain results equivalent to those from an sCMOS camera, yet achieve a reduction in file size of up to two orders of magnitude. This is important in the modern computer era, as it reduces data throughput, and is well-aligned with edge-computing applications. We believe the EBC is an excellent choice, particularly for long-term studies of active matter.
\end{abstract}

\section{Introduction}\label{sec:Introduction}

    Studies of both passive and active micro-particles have become increasingly prominent due to their diverse applications across many fields including bioscience, health care, and environmental sustainability~\cite{abdelmohsen_micro-_2014,balasubramanian_micromachine-enabled_2011,katuri_designing_2017,gao_environmental_2014,qu_nanotechnology_2013,karn_nanotechnology_2009}. Passive micro-particles are inert and exhibit Brownian motion as a result of collisions with thermal molecules in their environment. Understanding the trajectories of individual particles and their dynamics is of key importance to many applications, such as molecular transport~\cite{manzo_review_2015}, and understanding brain water diffusion~\cite{cherstvy_anomalous_2021, beaulieu_basis_2002}, where different regimes arise and diffusion can be non-ergodic. In contrast, active particles perform directed motion and drive their environment away from equilibrium~\cite{bechinger_active_2016}. Active particles have a wide range of applications: they have been used to power microscopic gears, micro-machines~\cite{singh_sperm_2020, sokolov_swimming_2010, vizsnyiczai_light_2017,di_leonardo_bacterial_2010}, and may underpin the next-generation platforms for minimally invasive medicine~\cite{patra_intelligent_2013, nelson_microrobots_2010}. Experimental studies of passive and active matter are inherently reliant on video microscopy. They depend on the successful identification and tracking of individual particles in complex environments over extended periods. In itself, this can present challenges for the imaging platforms employed to discern particle dynamics as well as the camera architecture used to extract the required particle trajectories.     
    
    Frame-based cameras are widely used in tracking studies. These include EMCCD (Electron Multiplying Charge Coupled Device) and sCMOS (scientific Complementary Metal Oxide Semiconductor)~\cite{peterkovic_optimising_2024}. These devices sample the pixel array synchronously, and produce images at a specified frame rate. Each image is comprised of the power detected per pixel integrated over the exposure time. This form of data acquisition may limit the investigation of passive and active matter dynamics. In particular, observing these processes often requires large fields of view, thereby restricting acquisition rates to ${\sim} 100$\,Hz and constraining the observable dynamic processes. Although high-speed cameras can overcome this to some degree, there is a trade-off: data acquisition rates may exceed GB.\,s$^{-1}$ which complicates data management and analysis~\cite{golibrzuch_application_2022}. 
    
    Recently, the neuromorphic, or event-based camera (EBC), has gained attention due to its unique functionality. Namely, the EBC samples its pixels asynchronously and detects changes in intensity. These characteristics potentially offer key advantages for a range of video microscopy studies, such as increased temporal resolution, reduced file sizes, and reduced power consumption.
    They have been used in applications including single-molecule localisation microscopy~\cite{cabriel_event-based_2023}, Fourier light field microscopy~\cite{guo_eventlfm_2024}, observation of levitated micro-particles~\cite{ren_event-based_2022,ren_neuromorphic_2024}, real-time velocity-resolved ion kinetics~\cite{golibrzuch_application_2022}, observation of plankton behaviour in situ~\cite{takatsuka_millisecond-scale_2024}, and automotive vision~\cite{gehrig_low-latency_2024}.
    
    In this paper, we focus on an experimental validation of the use of an EBC in both passive and active matter studies. To achieve this, we obtained side-by-side recordings of passive and active matter samples with an EBC and the more widely used sCMOS camera, allowing a detailed comparison of both approaches for the first time. We extracted key parameters of motion from both recordings and compared them. This allowed us to benchmark the EBC against the sCMOS platform. Firstly, we explored the case for passive matter: we observed the two-dimensional Brownian motion of micro-particles suspended in heavy water and determined their diffusion coefficient. We then turned to a study of active matter, where we analysed motility parameters of murine spermatozoa. Although initial work with an EBC and spermatozoa has recently been conducted~\cite{sadak_human_2024}, no studies comparing or validating the EBC against established imaging technologies exist in this context. This study represents the first direct comparison of parameters extracted from EBC datasets with those obtained from sCMOS datasets which were recorded simultaneously. Using the EBC, we were able to extract parameters of motion for both passive and active matter that agreed with those acquired with an sCMOS platform. This agreement persisted even when the EBC was tuned to report only large intensity changes, yielding a reduction in file sizes of up to two orders of magnitude (${\sim}155\text{-fold}$ for passive matter and ${\sim}418\text{-fold}$ for active matter) when compared to data acquired by an sCMOS camera. This major benefit of the EBC will be central to future studies of active matter, particularly those that require extended observation periods.

\section{Camera Architecture}\label{sec:CameraArchitecture}

\begin{figure}[htpb]
	   \centering
	   \includegraphics[width=0.9\linewidth]{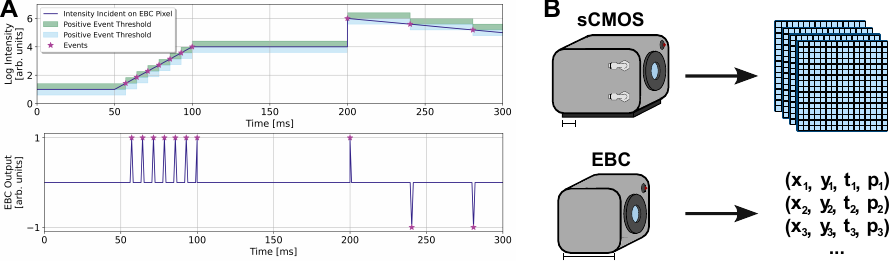}
	   \caption{\textbf{A.} Example log-intensity incident on a single EBC pixel, with shaded areas representing detection thresholds. Events are registered when the log-intensity increases or decreases past the threshold value. After an event is registered, the reference value of the pixel is updated to the log-intensity which prompted the event. The second plot shows the output from an EBC pixel for the input log-intensity signal given. Reported events are shown by stars. \textbf{B.} Illustration of difference in output formats between frame-based cameras (sCMOS) and EBCs. The sCMOS outputs a series of frames at a fixed rate, while the EBC outputs a list of asynchronous events. The scale bar is $3\,\text{cm}$ in both schematics.} 
	   \label{fig:EventBasedCameraBehaviour}
    \end{figure}

    Event-based cameras operate in a fundamentally different way to frame-based cameras, which leads to their unique functionality and a number of key advantages. In contrast to frame-based cameras, the EBC asynchronously detects changes in the logarithm of the detected intensity at a pixel, which is often termed `log-intensity'~\cite{gallego_event-based_2022}. The behaviour of a single EBC pixel for a given log-intensity time series is depicted in \autoref{fig:EventBasedCameraBehaviour}A. The asynchronicity of the pixel sampling allows for microsecond timestamp resolution and sub-millisecond readout latency~\cite{gallego_event-based_2022}. Instead of returning a series of images, the EBC output is a list, where each line corresponds to one such change in intensity, or ``event''. This is depicted in \autoref{fig:EventBasedCameraBehaviour}B. Each such event is characterised by the coordinates of the pixel $(x_i,\, y_i)$, the time at which the event was observed $t_i$, and the event polarity $p_i$: whether the intensity increased or decreased past a given threshold. This event data can be transformed into various formats, such as video, to enable visualisation and frame-based analysis techniques. There also exist analysis algorithms which may be applied directly to the native data~\cite{gallego_event-based_2022}.

    An advantage of the EBC is the absence of a fixed sampling rate, which allows for the exploration of dynamic processes without \textit{a priori} knowledge of their time scales~\cite{cabriel_event-based_2023}. Moreover, because the camera only records changes in intensity, stationary objects are not sampled and do not generate data. This makes data acquisition more efficient and allows longer-duration studies while maintaining high temporal resolution~\cite{golibrzuch_application_2022}. The EBC also consumes significantly less power than a typical sCMOS camera ($0.5\,\text{W}$ versus $100\,\text{W}$ in our case) and can be powered by a computer, facilitating integration into more complex systems or even portable setups~\cite{gallego_event-based_2022}. However, the functionality of the EBC may add a level of complexity to some aspects of these studies. In particular, the camera records the polarity of an event in a binary manner (i.e. increase or decrease in intensity), without reporting the exact intensity change.
    Additionally, moving objects generate leading and fading edges as they leave the purview of some pixels and enter new ones. Both of these factors present challenges for interpreting and analysing results and merit careful comparison with the sCMOS platform to ensure that both approaches yield equivalent results.
    
\begin{figure}[htbp]
	   \centering
	   \includegraphics[width=0.6\linewidth]{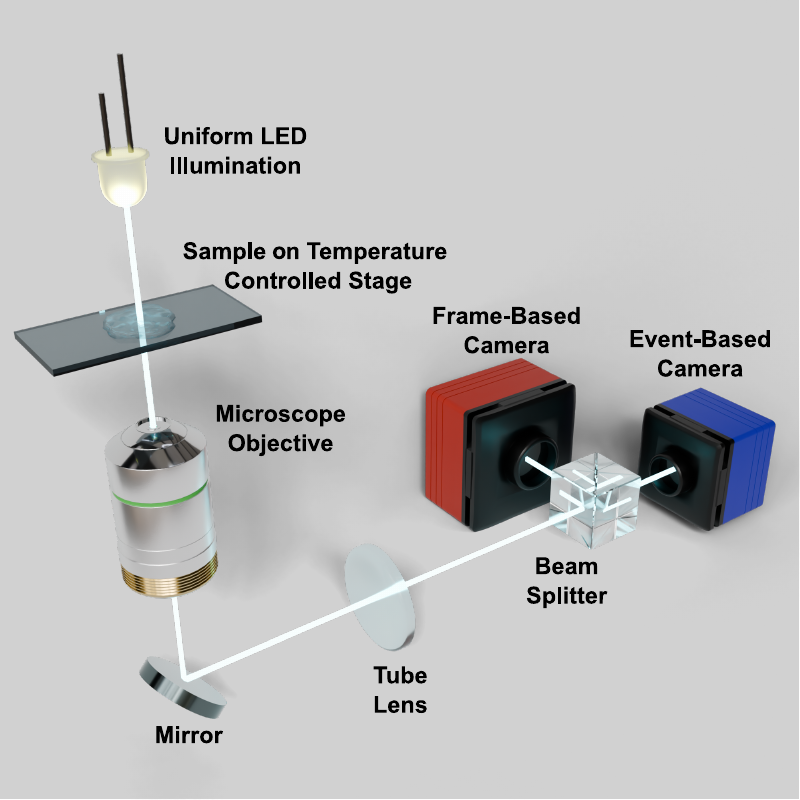}
	   \caption{Bespoke brightfield microscope used for imaging. The samples were contained within a temperature-controlled stage and kept at a near-constant temperature $(\pm 0.06\, ^\circ\text{C})$. K\"{o}hler illumination was implemented to uniformly illuminate the sample (lenses used not shown). A $300\,\text{mm}$ tube lens was used to image the sample plane and the 50-50 beam splitter, allowed for simultaneous measurements on the event-based and frame-based (sCMOS) cameras.} 
	   \label{fig:experimentalSetup}
    \end{figure}

    To be able to quantify the advantages and challenges of the EBC compared to an sCMOS camera, we use both to simultaneously image passive and active matter samples using a bespoke brightfield microscope shown in \autoref{fig:experimentalSetup}. However, each camera possessed a different field of view (FOV) due to their different pixel, and sensor array sizes: $6.5\, \text{\textmu m}$ pixels in a $2048\times 2048$ array for the sCMOS and $4.86\, \text{\textmu m}$ pixels in a $1280\times 720$ array for the EBC. To mitigate this and aid an appropriate comparison, the region of interest (ROI) on the larger sCMOS array was reduced to $956 \times 538$ pixels, which yielded an approximately equal FOV for both cameras.

\section{Passive Matter}\label{sec:PassiveMatter}

    Our investigation of passive matter served as a precursor to exploring active matter as it is well studied and understood. Thus, it was a good test for the EBC. We performed a comparison between the EBC and sCMOS by extracting the diffusion coefficients of different-sized micro-particles suspended in $\text{D}_2\text{O}$. Micro-particles, or beads, were chosen because of their prevalence as calibration tools for different imaging systems~\cite{valli_super-resolution_2021}. 
    
    We generated samples by progressively diluting the manufacturer's bead suspension in heavy water until only a few $(1-5)$ beads were visible in the FOV of the microscope. A total of $7.5\,\text{\textmu L}$ of the dilution was pipetted into the well of a $7.63\,\text{\textmu L}$ adhesive imaging spacer on a glass coverslip. Another glass coverslip was used to seal the sample in the spacer volume to prevent evaporation. The beads were imaged using a $40\times$ $0.65\,\text{NA}$ plan achromat microscope objective. A total magnification of $66.5\times$ was calculated by calibrating the system with a 1951 USAF resolution test target. During imaging, samples were kept at $37\, ^\circ\text{C}$ using a custom-built temperature-controlled stage (TCS). The TCS was heated using a ceramic ring heater controlled by an Arduino UNO. The temperature of the stage was monitored by a platinum resistance temperature detector which provided feedback for the heating. We calibrated the stage by recording the temperature inside a water-filled confocal dish in the sample holder. After 30 minutes, which allowed the system to come to equilibrium, the water temperature remained constant $(\pm 0.06\, ^\circ\text{C})$ for 50 minutes. 
    %of equilibration,
    
    The beads were imaged for 10 minutes on both cameras: the EBC recorded with its default settings and the sCMOS recorded with the reduced ROI at $20$ fps. These recordings were then divided into twelve $50$-second clips for averaging. Bead trajectories were extracted from the sCMOS and EBC datasets via centroid tracking. The sCMOS frames were analysed using TrackMate~\cite{ershov_trackmate_2022, tinevez_trackmate_2017}, an open-source Fiji (ImageJ) plugin. The EBC data was tracked using the general tracker provided by the EBC manufacturer in their Metavision SDK.
        
    The motion of colloidal Brownian particles may be characterised with a diffusion coefficient given by $D = k_B T/\gamma$, where $k_B$ is the Boltzmann constant, $T$ is the sample temperature, and $\gamma$ is the translational drag coefficient. For free diffusion of a spherical particle, setting $\gamma = 6\pi\eta a$ results in the well-known Stokes-Einstein-Sutherland equation, where $\eta$ and $a$ correspond to the suspension fluid viscosity and particle radius respectively~\cite{dhont_introduction_1996}. The mean-squared displacement (MSD) of the beads is related to the diffusion coefficient, and is given by:
    \begin{equation} 
       \langle r^2 \rangle_n  = \frac{1}{N-n}\sum_{i=1}^{N-n} (\textbf{\textit{r}}_{i+n} - \textbf{\textit{r}}_{i})^2 = 4D(n\Delta t)
        \label{eq:MSD}
    \end{equation}
    where $\textbf{\textit{r}}_{i+n}$ and $\textbf{\textit{r}}_{i}$ are the positions of a bead, measured in the $(i+n)^{\text{th}}$, and $i^{\text{th}}$ frames respectively. Physically, $n$ is the frame gap over which the square displacement is calculated, $N$ is the total number of frames, and $\Delta t$ refers to the time step between frames.
    
    To determine $D$, the mean-squared displacement is calculated for each $\tau=n\Delta t$ and a least-squares fit of the form $\langle r^2\rangle_\tau = a\tau + b $ is used. $D$ is then related to the fit parameters via \autoref{eq:MSD}. This method possesses fundamental limitations which have been addressed in numerous publications~\cite{qian_single_1991, michalet_mean_2010, michalet_optimal_2012}. Most pertinent to our work is the problem arising from the use of finite-sized trajectories: MSD values corresponding to larger values of $\tau$ are less averaged and consequently less accurate, thus necessitating longer trajectories for appropriately averaged squared displacements~\cite{ernst_measuring_2013, michalet_optimal_2012}. This effect is alleviated by fitting to an optimal number of points on the MSD vs $\tau$ curve~\cite{michalet_optimal_2012}. To further minimise the effect, we allowed the beads to sediment until they settled very close to the bottom coverslip. This ensured that they were near the bottom of their suspension volume and would not diffuse or sediment out of the focal plane of the microscope. This enabled the collection of a significant amount of data for the bead trajectories, increasing the accuracy of the measured diffusion coefficient.

    \begin{table}[htbp]
        \centering 
        \begin{tabular}{|c|cc|c|c|}
            \hline
            \multirow{3}{*}{\textbf{\begin{tabular}[c]{@{}c@{}}Bead \\ Diameter \boldmath$(2a)$\end{tabular}}} & \multicolumn{2}{c|}{\textbf{Diffusion Coefficient \boldmath$(D)$ \boldmath$\left[\textbf{\textmu} \text{m}^2\,\text{s}^{-1}\right]$}} & \multirow{3}{*}{\textbf{\begin{tabular}[c]{@{}c@{}} Expected \\ \boldmath$D$ Range for \\ \boldmath$a< s < 2a$ \boldmath$\left[\textbf{\textmu} \text{m}^2 \,\text{s}^{-1}\right]$\end{tabular}}} & \multirow{3}{*}{\textbf{\begin{tabular}[c]{@{}c@{}}\% Difference \\ Between \\ Cameras\end{tabular}}} \\ \cline{2-3}
                                                                                           & \multicolumn{1}{c|}{\multirow{2}{*}{\textbf{sCMOS}}}     & \multirow{2}{*}{\textbf{EBC}}     &                                                                                                                                                                                                                                      &                                                                                                       \\
                                                                                           & \multicolumn{1}{c|}{}                                    &                                   &                                                                                                                                                                                                                                      &                                                                                                       \\ \hline
            % $0.7\,\text{\textmu m}$                                                                   & \multicolumn{1}{c|}{$0.502 \pm 0.037$}                 & $0.444 \pm 0.147$               & $0.436 - 0.570$                                                                                                                                                                                                                    & $12.34$                                                                                               \\ \hline
            % $1.0 \, \text{\textmu m}$                                                                  & \multicolumn{1}{c|}{$0.293 \pm 0.014$}                 & $0.324 \pm 0.021$               & $0.036 - 0.383$                                                                                                                                                                                                                    & $10.13$                                                                                               \\ \hline
            $1.6\, \text{\textmu m}$                                                                   & \multicolumn{1}{c|}{$0.206 \pm 0.031$}                 & $0.214 \pm 0.029$               & $0.022 - 0.240$                                                                                                                                                                                                                      & $4.26$                                                                                               \\ \hline
            $3.2\, \text{\textmu m}$                                                                   & \multicolumn{1}{c|}{$0.067 \pm 0.004$}                 & $0.068 \pm 0.008$               & $0.015 - 0.120$                                                                                                                                                                                                                    & $0.89$                                                                                               \\ \hline
        \end{tabular}
        \caption{Mean diffusion coefficients, $D$, extracted from sCMOS and EBC datasets utilising the optimised least-square fit (OLSF) from Ref.~\cite{michalet_optimal_2012}. The corrected range for $D$ is obtained by considering the expected diffusion coefficient when the bead centre is between one or two radii from the glass slide~\cite{carbajal-tinoco_asymmetry_2007}. Values are reported as mean $\pm$ standard deviation.}
        \label{tab:DiffCoeffResults}
    \end{table}
    
    \autoref{tab:DiffCoeffResults} summarises the diffusion coefficients we extracted from the sCMOS and EBC data, showing they are within the expected ranges, and that the cameras agreed well with each other. Percentage differences between the cameras larger than $5\%$ were observed for micro-particles smaller than $1\,\text{\textmu m}$. In these cases, particles would diffuse out of the focal plane of the microscope, confounding the tracking for data taken on both cameras. We present a range of expected diffusion coefficients because the drag experienced by a particle increases near a confining surface (such as the coverslip) as a consequence of the boundary conditions imposed on the hydrodynamics~\cite{carbajal-tinoco_asymmetry_2007}. The magnitude of the decrease in $D$ is dependent on the proximity of the particle to the surface, denoted as $s$ in the table. As we were not able to accurately measure and/or fix this distance, we provide the range of diffusion coefficients expected for distance variations of the scale of one particle radius $(a)$.
    
\section{Active Matter}\label{sec:ActiveMatter}
    For the study with active matter, we used murine spermatozoa and extracted basic motility parameters from the trajectories of the cells. In fertility clinics, semen analysis is a fundamental aspect of male fertility assessment and sperm motility evaluation is an important feature of this analysis~\cite{boitrelle_sixth_2021}. While manual assessment by an expert is considered the gold standard by the World Health Organisation (WHO)~\cite{world_health_organisation_laboratory_2021}, it is costly and time-consuming. An alternative is computer-aided sperm analysis (CASA), which has been applied to agriculturally important animals, such as horses, pigs, and rams, to assess the quality of spermatozoa for optimising insemination and improving reproductive outcomes~\cite{van_der_horst_computer_2020}. CASA is commonly used in research, and increasingly in commercial and clinical settings. It relies on tracking spermatozoa in two or three dimensions, which can generate very large files and limit the duration of studies. 
    %This makes reducing data throughput without loss of information a pertinent area for exploration and suggests that analysis may benefit from the particular advantages of the EBC.
    This makes reducing data throughput without loss of information a pertinent area for exploration. Using EBCs in these applications will allow for longer studies and may benefit analysis.
   
   Our goals were to demonstrate the applicability of the EBC to sperm motility studies, to utilise the unique sampling properties of this camera, and to analyse the impact on data throughput. In particular, the high temporal resolution of the EBC can only be exploited by objects which move across pixels quickly, such as spermatozoa. This arises from the distinction between the pixel dead time, and the temporal resolution of the events it records, which in our case are approximately $1\,\text{ms}$ and, $1\,\text{\textmu s}$ respectively.

    \begin{figure}[htbp]
	   \centering
	   \includegraphics[width=0.9\linewidth]{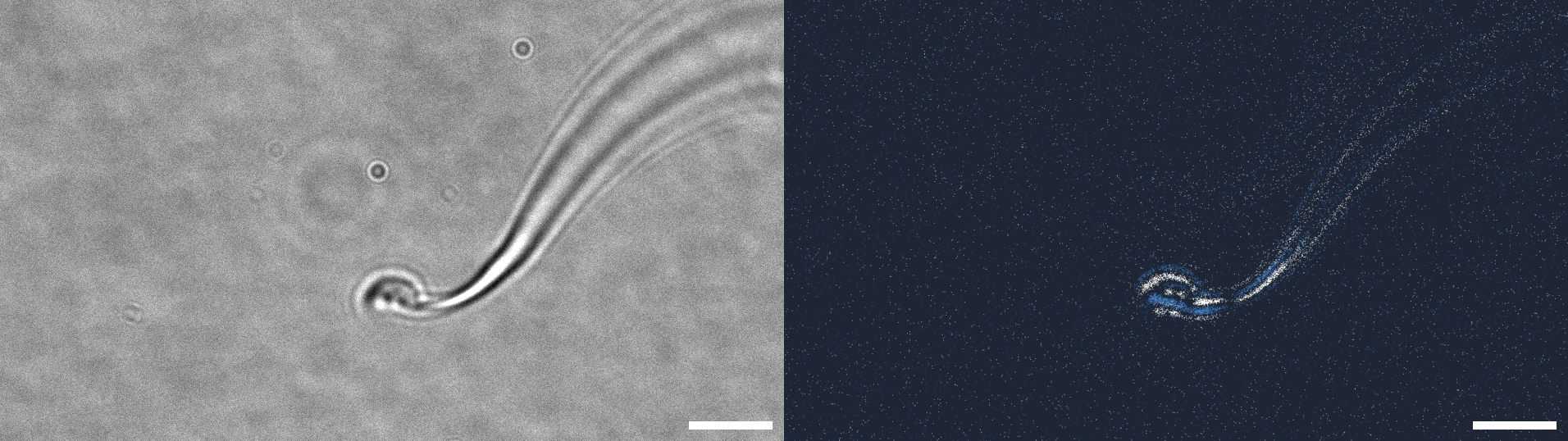}
	   \caption{Comparison of a murine spermatozoon imaged by the sCMOS camera (left) and EBC (right) after we have converted the EBC data to video for visualisation. The dark background on the EBC image corresponds to no events, while the white and blue pixels correspond to events with positive and negative polarity respectively. The single spermatozoon imaged was within a sample diluted to a final concentration of $1 \%$ spermatozoa. The scale bar is $10\,\text{\textmu m}$ in both images.} 
	   \label{fig:spermComparison}
    \end{figure}
      
    All animal studies associated with this project were conducted in accordance with the Australian Code of Practice for the Care and Use of Animals for Scientific Purposes. To harvest spermatozoa samples, male mice of proven fertility were culled by cervical dislocation with the epididymis and vas deferens collected in Research Wash Medium (ART Lab Solutions) supplemented with 4 mg $\text{ml}^{-1}$ bovine serum albumin (BSA). Spermatozoa were released from the vas deferens and caudal regions of the epididymis by blunt dissection in $1\,\text{mL}$ of Research Fertilisation Medium (ART Lab Solutions) supplemented with 4 mg $\text{ml}^{-1}$ BSA, and allowed to capacitate for $1$ hour in an incubator with a humidified atmosphere of 5\% $\text{O}_2$, 6\% $\text{CO}_2$, and a balance of $\text{N}_2$ at $37\, ^\circ\text{C}$. Following capacitation, spermatozoa were counted using a haemocytometer and the concentration was determined within a range of $1-3 \times 10^6$ cells per mL . For imaging, the sample was then diluted in Research Fertilisation Medium to achieve a final concentration of $1 \%$ within a $2\,\text{\textmu L}$ droplet of Research Wash Medium supplemented with 4 mg $\text{ml}^{-1}$ BSA in a confocal dish (Ibidi, Martinsried, Planegg, Germany). To minimise evaporation, droplets were overlaid with paraffin oil.

    %Spermatozoa were then transferred into $2\,\text{\textmu L}$ drops of Research Wash medium in a pre-equilibrated confocal dish (Ibidi, Martinsried, Planegg, Germany) overlaid with paraffin oil to prevent evaporation. (Meg: this can be removed)
    
    Samples were imaged in the brightfield microscope depicted in \autoref{fig:experimentalSetup} with the temperature of the confocal dish maintained at $37\, ^\circ\text{C}$ using a heated temperature-controlled-stage. A $10\times$ $0.25\,\text{NA}$ plan achromat microscope objective was used instead of the $40\times$ to account for the size of the sperm cell heads (${\sim}6.5-8.5\,\text{\textmu m}$~\cite{vareasanchezGeometricMorphometricsRodent2013}). The system was re-calibrated using the same 1951 USAF resolution test target, yielding a total magnification of $16.7\times$. The sCMOS camera recorded data at $57\, \text{fps}$, it also utilised the same reduced ROI as described above. As before, the EBC recorded data using default settings. \autoref{fig:spermComparison} shows a murine spermatozoon imaged in the system after the EBC data has been converted to video for visualisation. Similar to the passive matter study, both TrackMate and the Metavision SDK were used to perform tracking of the cells in the sCMOS and EBC datasets respectively. 
        
\begin{figure}[htbp]
	   \centering
	   \includegraphics[width=0.8\linewidth]{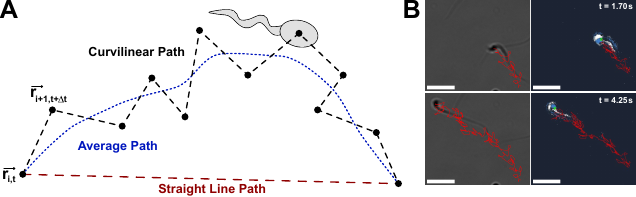}
	   \caption{\textbf{A.} Schematic of a spermatozoon swimming adapted from~\cite{amann_computer-assisted_2014}. The black circles represent the position of a spermatozoon in different video frames (separated in time by $\Delta t$). The black path indicates the motion of the cell from one frame to another. The blue path shows a roaming average of the position of the cell. Finally, the red path connects the first and the last point where the cell was detected. \textbf{B.} Images of the trajectory of one spermatozoon at two different times in the same recording, to illustrate tracking. The sCMOS image is shown on the left and data from the EBC is shown on the right after it has been processed into an image for visualisation. The scale bar is $30\,\text{\textmu m}$.}
	   \label{fig:SpermMotilityParameters}
    \end{figure}

    Three parameters are commonly used when analysing the motility of spermatozoa: curvilinear velocity (VCL), average path velocity (VAP), and straight-line velocity (VSL). VCL measures the mean velocity of the spermatozoon from one detected position to another~\cite{world_health_organisation_laboratory_2021}. VAP refers to the mean velocity along a path that is a roaming average of the position of the cell over a given number of frames~\cite{wilson-leedy_development_2007}. Finally, VSL is the velocity along the path from the first detected position to the last~\cite{world_health_organisation_laboratory_2021}. \autoref{fig:SpermMotilityParameters} shows a schematic of the paths along which these velocities are calculated as well as an example of the trajectory and track of a single spermatozoon at different times. It is worth noting that despite being referred to as velocities, these values are calculated using distances, not displacements. CASA systems often measure numerous other variables; we restricted our comparison to these three, as most of the other parameters are derived from these measurements~\cite{world_health_organisation_laboratory_2021}.

    The VCL, VAP, and, VSL velocity parameters are easily determined from the tracked positions of a spermatozoon. VCL and VSL are given by the following equations~\cite{alquezar-baeta_opencasa_2019}:
     \begin{eqnarray}
        \text{VCL} & = & \frac{1}{N-1}\sum_{i=1}^{N} \frac{\left|\textbf{\textit{r}}_i -\textbf{\textit{r}}_{i-1}\right|}{t_{i} - t_{i-1}}  \label{eq:VCL} \\
        \text{VSL} & = & \frac{\left|\textbf{\textit{r}}_{N} -\textbf{\textit{r}}_{0}\right|}{t_{N} - t_{0}}
        \label{eq:VSL}
    \end{eqnarray}
    where $N$ is the total number of frames, $\textbf{\textit{r}}$ refers to the tracked position of the cell in a frame, $t$ refers to the time of a frame, and the subscripts $i$, $(i-1)$, $ N$, and $0$ correspond to the specific frames; with $0$ and $N$ being the first and last respectively. Determining VAP first requires finding an average path via ${\textbf{\textit{q}}_{i} = \frac{1}{w}\sum_{j=i}^{i+w} \textbf{\textit{r}}_j}$. This corresponds to a roaming average of the tracked positions of the cell over a window of $w$ frames. This average path $\textbf{\textit{q}}_i$ is then used to determine VAP~\cite{alquezar-baeta_opencasa_2019}:
    \begin{equation}
        \text{VAP} = \frac{1}{N-w}\sum_{i=1}^{N-w} \frac{\left|\textbf{\textit{q}}_i -\textbf{\textit{q}}_{i-1}\right|}{t_{i} - t_{i-1}}
        \label{eq:VAP}
    \end{equation}

    \autoref{tab:prelimMotilityResults} shows a set of velocity parameters obtained from a recording of one spermatozoon travelling across the FOV of both cameras.
    For this particular recording, the cell was in the FOV of the microscope for a total 277 frames, or approximately 4.8 seconds. 
    The frame-to-frame velocity was calculated giving 277 values with the VCL, calculated as per \autoref{eq:VCL}, and standard deviation is reported in \autoref{tab:prelimMotilityResults}. 
    %The frame-to-frame velocities were then used to obtain a standard deviation. 
    A roaming average window of $w=10$ frames was used to determine the averaged path for calculating the VAP, yielding an average path containing 267 points. 
    The velocity between each point on the averaged path was calculated with the mean VAP,  calculated per \autoref{eq:VAP}, and standard deviation reported in \autoref{tab:prelimMotilityResults}. 
    Finally, the VSL was calculated using \autoref{eq:VSL}. Since this value is not averaged, we do not report a standard deviation. 
    The values reported in \autoref{tab:prelimMotilityResults} agree well with literature values~\cite{van_der_horst_computer_2020, goodson_classification_2011}.
    
    \begin{table}[h!]
        \centering
        \begin{tabular}{|c|c|c|c|}
        \hline
        \textbf{\begin{tabular}[c]{@{}c@{}}Parameter\\ \boldmath$[\text{\textmu m}\,\text{s}^{-1}]$\end{tabular}} & \multicolumn{1}{c|}{\textbf{sCMOS}} & \multicolumn{1}{c|}{\textbf{EBC}} & \multicolumn{1}{c|}{\textbf{\begin{tabular}[c]{@{}c@{}}Difference Between \\ Cameras \boldmath$[\text{\textmu m}\,\text{s}^{-1}]$\end{tabular}}} \\ \hline
        \textbf{VCL} &  $152.46\pm 98.27$ & $157.81\pm 99.04$ &   $5.35$ \\
        \hline
        \textbf{VAP} &  $41.89\pm 21.37$ &  $39.31 \pm 21.86$ & $2.58$  \\
        \hline
        \textbf{VSL} &  $25.30$ &           $21.59$ &           $3.71$  \\ 
        \hline
        \end{tabular}
        \caption{Velocity values obtained from the recording of the spermatozoon in \autoref{fig:SpermMotilityParameters}, swimming across the FOV of both cameras. Values are reported as mean $\pm$ standard deviation, except for VSL which reports a single value for a single recording.}
        \label{tab:prelimMotilityResults}
    \end{table}

    We observed excellent agreement between the cameras for all observed parameters. It is worthwhile noting that spermatozoa are heterogeneous and do not swim with constant velocity. Consequently, the standard deviations in \autoref{tab:prelimMotilityResults} are better interpreted as characteristics of a distribution of velocities of the spermatozoon, rather than errors, or uncertainties, in the velocity parameters. \autoref{tab:spermPercentDiffs} shows the percentage differences between the EBC and sCMOS for velocity parameters obtained from four different sperm cells. The curvilinear velocity (VCL) parameter yielded the worst agreement between the cameras. This is due to spermatozoa swimming slightly in and out of focus, confounding the cell tracking for both cameras. This meant that the appropriate centroid position was more difficult to determine frame-by-frame. This effect is not as pronounced for the VAP as the roaming average used to generate the path reduces the uncertainty on the centroid position. Similarly, the VSL value is not significantly affected as the uncertainty on the initial and final centroids is small compared to the distance the cell has travelled.
    %Great paragraph Gabe!
    
    \begin{table}[h!]
        \centering
        \begin{tabular}{|c|c|}
            \hline
            \textbf{Parameter} & \textbf{\% Differences Between Cameras} \\ 
            \hline
            \textbf{VCL} & $13.53\pm9.44$ \\ 
            \hline
            \textbf{VAP} & $4.85\pm2.22$ \\ 
            \hline
            \textbf{VSL} & $7.50\pm5.70$    \\
            \hline
        \end{tabular}
        \caption{Differences of the velocity parameters extracted from four different spermatozoa from four different recordings, each recorded simultaneously by the EBC and sCMOS cameras. Values are reported as mean $\pm$ standard deviation.}
        \label{tab:spermPercentDiffs}
    \end{table}

\section{File Sizes}\label{sec:FileSizes}

    A key motivation for applying EBCs to passive and active matter studies was the reduction of redundant data and its consequent impact on file sizes. Recent publications cite very large decreases in data throughput when switching from a frame-based camera to an EBC in different applications, claiming reductions as large as $1200$-fold~\cite{golibrzuch_application_2022, ren_neuromorphic_2024, takatsuka_millisecond-scale_2024}. However, the parameters used to obtain these large reductions are not explicit in some cases, complicating direct comparisons between the cameras. 
    
    Using the experimental parameters described previously, we achieved a $5\text{-fold}$ reduction in raw data size for beads and $3\text{-fold}$ for spermatozoa. However, these reductions are highly dependent on system parameters. By engineering experimental parameters, we were able to demonstrate a reduction in file size of two orders of magnitude by: adjusting event contrast thresholds, using an equivalent number of pixels for comparison (instead of an equal FOV), increasing the sCMOS image bit depth to 16 bits, and increasing the brightness of the illumination LED. With these new parameters, we observed a maximum $155\text{-fold}$ reduction in file size while tracking a single $1.6\, \text{\textmu m}$ diameter bead. In estimating the diffusion coefficient of this micro-particle (as in \autoref{sec:PassiveMatter}), we observed an $8.5\%$ difference between the values obtained from the sCMOS and EBC data. Additionally, the estimated values from both cameras were also consistent with the corrected range for $D$ we presented in \autoref{tab:DiffCoeffResults}. In tracking a single murine spermatozoon with these engineered parameters, we observed a maximum $418\text{-fold}$ reduction in file size. It should be noted that the spermatozoa recordings for the EBC and sCMOS are each 1 minute long and that the spermatozoon is only in the FOV of the microscope for ${\sim}5$ seconds. If we only compare the period where the cell is visible in the FOV, then the file size reduction is more modest ($84\text{-fold}$).
    
    \begin{figure}[h]
        \centering
        \includegraphics[width=0.34\linewidth]{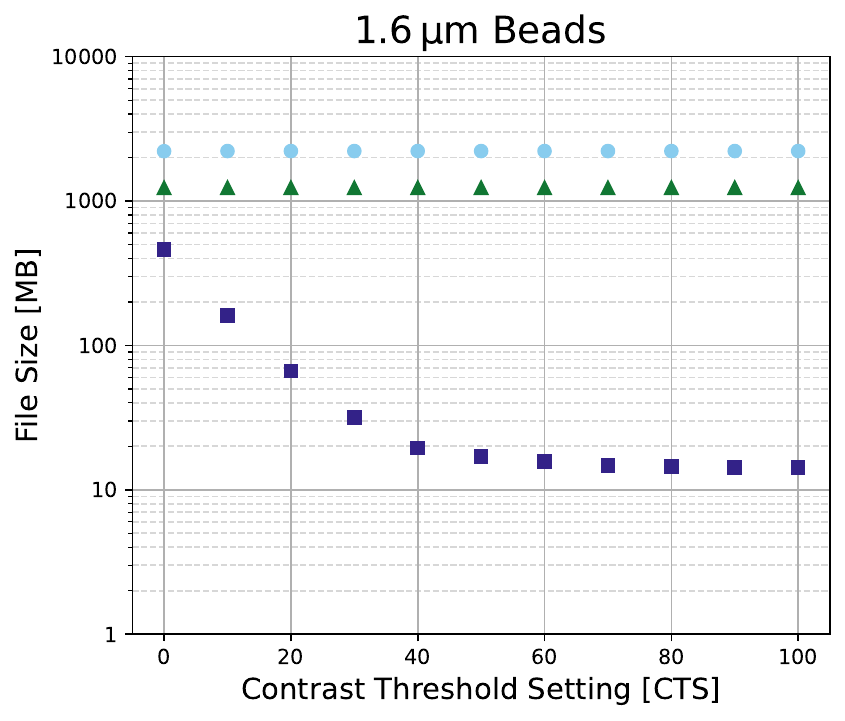}
        \includegraphics[width=0.34\linewidth]{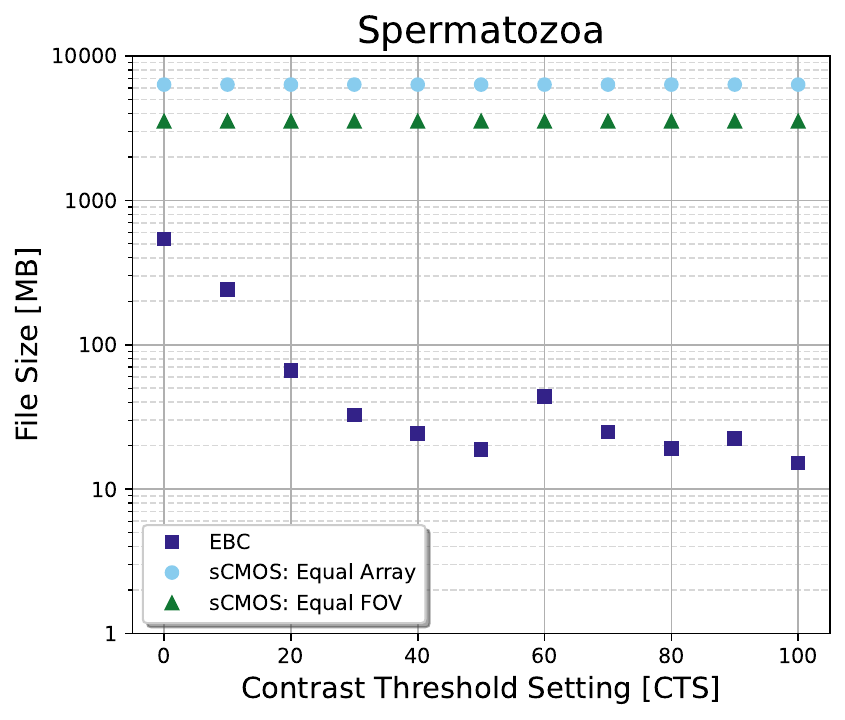}
        \caption{Output file size dependence on EBC contrast threshold setting (CTS). With no tuning (CTS$\,{=}\,0$), the file sizes between sCMOS and the EBC are comparable. However, as we increase the CTS, the file size advantage of the EBC becomes apparent. The peaks in the spermatozoa trend arise from varying numbers of spermatozoa swimming in and out of the FOV of the microscope, leading to more or less events correspondingly. The concentration of the bead and spermatozoa samples used for this data were $0.002\%$ and $1 \%$  respectively.}
        % Meg comment: Please relabel right graph to replace "Sperm" with "Spermatozoa"
        \label{fig:FileSizeThresholding}
    \end{figure}

    The event contrast threshold is particularly pertinent in relation to output file size. Increasing this parameter corresponds to requiring a larger log-intensity change for event detection, thereby decreasing the number of events and resultant file size. Separate parameters control the positive and negative event thresholds, but for simplicity, we adjusted them equally and referred to them collectively as the `contrast threshold setting' (CTS). To illustrate the impact of altering the CTS, we repeated measurements of the motion of $1.6\,\text{\textmu m}$ beads and murine spermatozoa in the brightfield microscope shown in \autoref{fig:experimentalSetup}. The sCMOS camera recorded with a bit depth of 16-bits and frame rates of 20\,fps for bead diffusion, and 57\,fps for sperm motility. The EBC recorded data with all default settings except for the modified CTS. Both cameras recorded data for one minute with the resultant file sizes summarised in \autoref{fig:FileSizeThresholding}. We show data for two different sCMOS file sizes, one corresponds to an array size equal to that of the EBC ($1280\times 720$), and the other with the reduced array size that yielded an equivalent FOV to the EBC ($956\times 538$). In both cases, the EBC file size was comparable to the sCMOS when there was no tuning of the CTS. We observe that as the CTS was increased, the number of events recorded by the EBC decreased, resulting in significantly smaller file sizes. The EBC file size plateaus for high CTS values as most noise events are eliminated, leaving only events generated by motile objects. This plateau is fairly smooth for beads. However, there are peaks in the spermatozoa file size trend which correspond to a varying number of spermatozoa swimming in the microscope FOV during recording.

    As the EBC responds to changes in brightness, recordings with more motile objects in them possess more events and correspondingly larger files. In fact, the file size advantage of the EBC only manifests when the object of interest is small and fast-moving in a large and constant background. This gives rise to a regime where the majority of pixels on the EBC are not sampled at any given time, thus yielding lower data throughput than a frame-based camera. If we consider the units of information for the sCMOS and EBC as the grayscale value from a single pixel and a single event respectively, then the EBC requires more data than the sCMOS per unit of information. In our case, the greyscale value from an sCMOS pixel was at most 16-bits. Contrasting this with our EBC where a single event timestamp was 24-bit, the coordinates of the pixel which observed the event were 11-bits for each $x$ and $y$, and event polarity was 1-bit, requiring a total of 47-bits for one event. If the intensities incident on all EBC pixels were to change simultaneously (e.g. by using a fast strobe light or by moving the camera through complex scenery), the frame-based cameras would produce smaller file sizes than the EBC. Consequently, the expected file size differences between the EBC and sCMOS are highly dependent on the application. This makes direct comparisons challenging and supports careful consideration of the best camera to use in a given instance.
    
    \begin{figure}[htbp]
        \centering
        \includegraphics[width=0.34\linewidth]{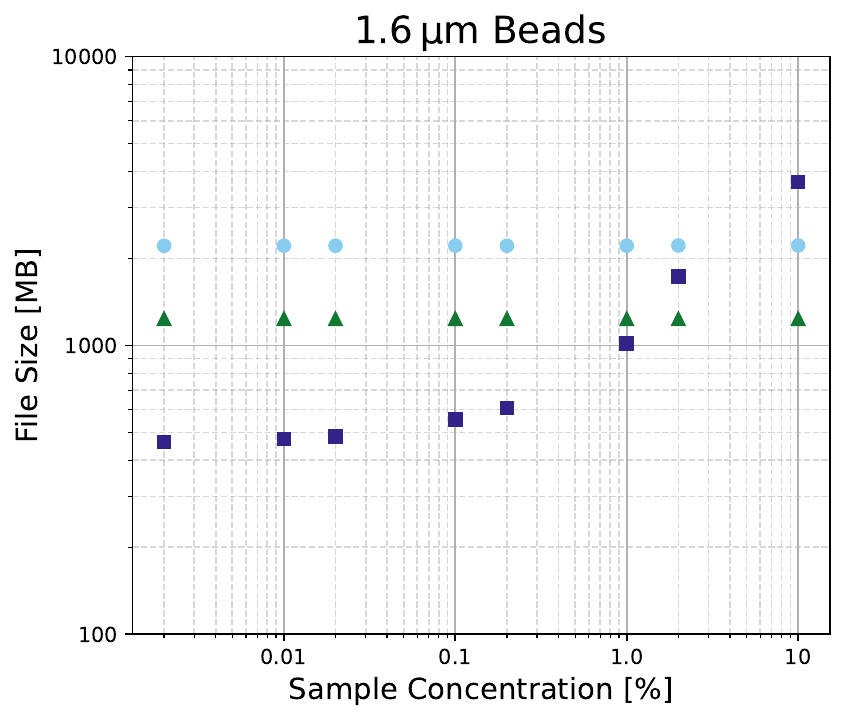}
        \includegraphics[width=0.34\linewidth]{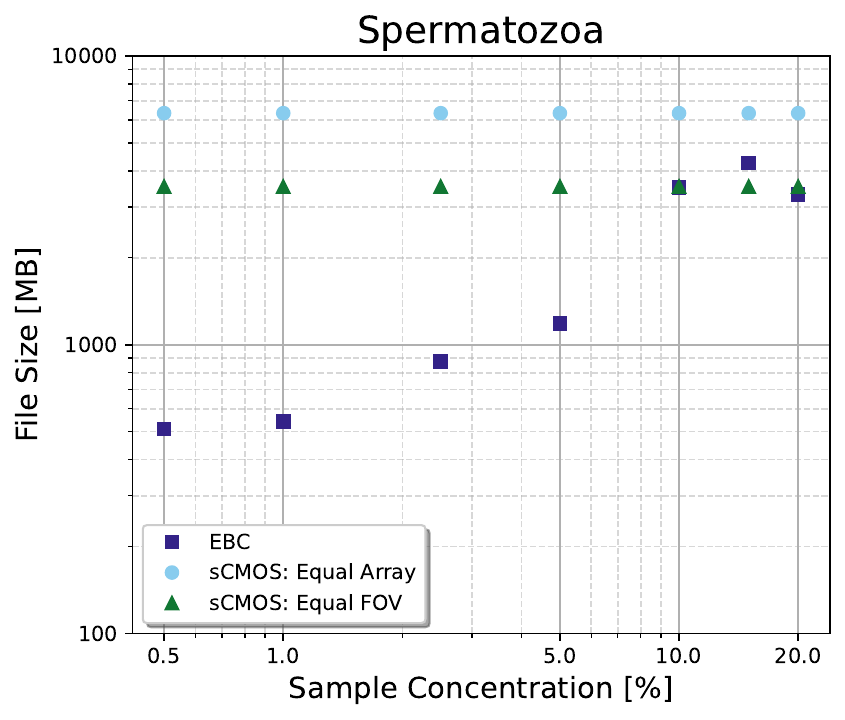}
        \caption{Output file size dependence on sample concentration. More concentrated samples possess a larger number of motile particles. This causes the EBC to register more events and record a larger file. As in \autoref{fig:FileSizeThresholding}, varying numbers of cells in the spermatozoa recordings introduces some variance to file sizes which is independent of the sample concentration. This explains why the largest file was obtained for $15\%$ concentration rather than $20\%$. It also explains the large drop off in file size between the $10\%$ and $5\%$ sample concentrations.
        The data presented here was recorded with the CTS set to its default of 0.}
        \label{fig:FileSizeConcentration}
    \end{figure}
    
    To explore this, we repeated the imaging of passive and active matter samples with varying concentrations. Higher concentrations exhibited a larger number of motile particles, which led to the sampling of more pixels. In turn, this yielded more events per unit time and larger data files. For passive matter, we created eight different dilutions from the manufacturer's concentrated $1.6\,\text{\textmu m}$ diameter micro-particle suspension. For the active matter, murine spermatozoa were harvested as described in \autoref{sec:ActiveMatter} and then further diluted in Research Fertilisation medium into seven samples with varying concentrations ($20\%$, $15\%$, $10\%$, $5\%$, $2.5\%$, $1\%$, and $0.5\%$). Both passive and active matter samples were imaged for 1 minute. The sCMOS recorded data as previously, and the EBC was tuned to its default settings. The output file sizes for the different concentrations of passive and active matter samples are shown in \autoref{fig:FileSizeConcentration}. 
    At low concentrations, both the beads and spermatozoa EBC file sizes plateau to a constant value which is a result of detecting noise events.
    As shown in \autoref{fig:FileSizeThresholding}, increasing the CTS can reject the detection of these noise events, further reducing the file size.
    As the concentration of the sample increases, so too does the file size as more events are recorded.
    It can even be seen that at high concentrations the EBC yields a larger file than the sCMOS camera.    

\section{Future perspectives for active matter studies}\label{sec:ImprovingSpermStudy}
    The work presented here shows that the EBC can be applied to studies of active matter. As mentioned earlier, tracking for both cameras worked optimally when spermatozoa were confined to a single imaging plane. Interestingly, human spermatozoa can exhibit a two-dimensional (2D) swimming mode when dispersed in a high viscosity medium and constrained close to a surface $(1\,\text{\textmu m})$: Conditions analogous to the female reproductive tract as spermatozoa ascend towards the ovulated oocyte~\cite{yazdan_parast_viscous_2024, yazdan_parast_cooperative_2023, nosrati_two-dimensional_2015}. Recapitulating these conditions in vitro would form an example of a study that could be undertaken. The EBC would enable the collection of larger datasets for verification, reduce the size of the files required for analysis, and facilitate long-duration studies. %As well as assisting in verification, studies in confined geometries possess biological relevance. For example, the lumen of the fallopian tubes possess curvature radii ranging from $20\,\text{\textmu m}$ to more than $150\,\text{\textmu m}$ ~\cite{raveshi_curvature_2021}. }$
    
    %These geometries lend themselves to use by the EBC for sperm motility studies, and active matter studies more broadly. 

    Utilisation of different illumination and imaging systems may further enhance the precision of tracking active particles with an EBC. For example, pulsed illumination is regularly used in some prominent super-resolution imaging methods for excitation. For single-molecule localisation microscopy using a pulsed source, the EBC was used to accurately resolve fluorescent molecules in a fluorophore-dense sample. On the other hand, a traditional frame-based camera exhibited limitations in the same scenario~\cite{cabriel_event-based_2023}.
    Pulsed illumination (using LEDs, amplitude-modulated lasers, or even femtosecond mode-locked lasers or frequency combs) in darkfield microscopes has also been utilised to scatter light off particles of interest~\cite{willert_event-based_2023, guo_eventlfm_2024} for advanced microscopy. This results in the whole particle being detected and imaged by the EBC, in contrast to the moving edge detection used in this study.
    Similar techniques could be employed in future to make particle tracking more robust and enhance the precision of tracking active particles.

    More broadly, neuromorphic cameras may find applications in edge-computing which aims to reduce the quantity of data sent over networks by processing data at the sensor, instead of remote processing in the cloud or at an appropriate server~\cite{cao_overview_2020}.
    There is a growing volume of research focused on addressing this problem and one goal for many applications is to perform image processing at the sensor level. This includes work on optical implementation of neural networks~\cite{lin_all-optical_2018, yan_fourier-space_2019, colburn_optical_2019} such as object classification~\cite{zheng_meta-optic_2022}, and object edge detection~\cite{qiu_spiral_2018, swartz_broadband_2024}.
    Object edge detection is akin to the functionality of the neuromorphic cameras used here, with the added advantage that EBCs only report events when an edge is in motion creating brightness changes, thereby rejecting all stationary background objects and reducing data transmission requirements.
    This is demonstrated here in our work by the smaller file sizes created by the EBC when compared to the sCMOS camera.
    Thus, the EBC may be employed in future edge-computing roles in various microscopy and manipulation applications~\cite{cabriel_event-based_2023, guo_eventlfm_2024, ren_event-based_2022, ren_neuromorphic_2024, takatsuka_millisecond-scale_2024}.
    
\section{Conclusion}\label{sec:Conclusion}
    Using a bespoke brightfield microscope we were able to record the Brownian motion of different sized microscopic beads on an event-based camera and a frame-based sCMOS camera simultaneously. By tracking the motion of the particles and analysing their trajectories we estimated their diffusion coefficients and observed agreement with expectation in all cases. Agreement between the cameras was observed within $<5\%$. Moreover, we employed the same microscope to observe murine spermatozoa motility on both the EBC and sCMOS cameras. We extracted basic motility parameters from spermatozoa tracks and again observed agreement (within $6\, \text{\textmu m}\,\text{s}^{-1}$) between event-based, and frame-based cameras. The unique sampling characteristics of the EBC are well suited to addressing problems in passive and active matter studies, where the dynamics of a small motile object are seen in a large FOV, in a sparse sample. This was evident when applying the EBC to tracking a single $1.6\,\text{\textmu m}$ bead, where we obtained a maximum $155\text{-fold}$ reduction in data file size when compared to sCMOS. We believe the EBC offers a particularly useful approach for cell motility studies which could benefit directly from the high temporal resolution and low data throughput of the EBC. Such qualities also make EBCs of interest to edge-computing applications.

\medskip
% \textbf{Supporting Information} \par %Please delete the Suppporting Information statement if it is not applicable. Please supply Supporting Information in another file. Supporting information should not be provided in .tex format
% Supporting Information is available from the Wiley Online Library or from the author.

% Acknowledgements
\medskip
\textbf{Acknowledgements.   }
The authors would like to thank Dr. Reza Nosrati for useful discussions. 
G. Britto Monteiro thanks An\'e Kritzinger, Emi Hughes, and Dr. Ralf Mouthaan from the Centre of Light for Life for insightful discussions. We acknowledge Alexander Trowbridge for assistance in generating the illustration in \autoref{fig:experimentalSetup}. This work was supported by the Australian Research Council (FL210100099), and the National Health and Medical Research Council (APP2003786). This research was supported by the Australian Research Council Centre of Excellence in Optical Microcombs for Breakthrough Science (project number CE230100006) and funded by the Australian Government.
Dr. Kylie Dunning is supported by a Future Making Fellowship (University of Adelaide).
This work used the NCRIS and Government of South Australia enabled Australian National Fabrication Facility - South Australian Node (ANFF-SA).
The University of Sydney authors acknowledge the NSW Smart Sensing Network (NSSN) sponsored by the NSW Office of Chief Scientist and Engineer as well as the Royal Australian Air Force that have supported aspects of the underpinning R\&D for this work.

\textbf{Data Availability.   }
The data that support the findings of this study are available from the corresponding author upon reasonable request.

% References
\medskip

% \bibliographystyle{unsrt}
% \bibliography{EBC_bibTeX.bib}

\begin{thebibliography}{10}

\bibitem{abdelmohsen_micro-_2014}
Loai K. E.~A. Abdelmohsen, Fei Peng, Yingfeng Tu, and Daniela~A. Wilson.
\newblock Micro- and nano-motors for biomedical applications.
\newblock {\em J. Mater. Chem. B}, 2:2395--2408, 2014.

\bibitem{balasubramanian_micromachine-enabled_2011}
Shankar Balasubramanian, Daniel Kagan, Che-Ming Jack~Hu, Susana Campuzano, M.~Jesus Lobo-Castañon, Nicole Lim, Dae~Y. Kang, Maria Zimmerman, Liangfang Zhang, and Joseph Wang.
\newblock Micromachine-{Enabled} {Capture} and {Isolation} of {Cancer} {Cells} in {Complex} {Media}.
\newblock {\em Angewandte Chemie International Edition}, 50:4161--4164, 2011.

\bibitem{katuri_designing_2017}
Jaideep Katuri, Xing Ma, Morgan~M. Stanton, and Samuel Sánchez.
\newblock Designing {Micro}- and {Nanoswimmers} for {Specific} {Applications}.
\newblock {\em Accounts of Chemical Research}, 50:2--11, 2017.

\bibitem{gao_environmental_2014}
Wei Gao and Joseph Wang.
\newblock The {Environmental} {Impact} of {Micro}/{Nanomachines}: {A} {Review}.
\newblock {\em ACS Nano}, 8:3170--3180, 2014.

\bibitem{qu_nanotechnology_2013}
Xiaolei Qu, Jonathon Brame, Qilin Li, and Pedro J.~J. Alvarez.
\newblock Nanotechnology for a {Safe} and {Sustainable} {Water} {Supply}: {Enabling} {Integrated} {Water} {Treatment} and {Reuse}.
\newblock {\em Accounts of Chemical Research}, 46:834--843, 2013.

\bibitem{karn_nanotechnology_2009}
Barbara Karn, Todd Kuiken, and Martha Otto.
\newblock Nanotechnology and in {Situ} {Remediation}: {A} {Review} of the {Benefits} and {Potential} {Risks}.
\newblock {\em Environmental Health Perspectives}, 117:1813--1831, 2009.

\bibitem{manzo_review_2015}
Carlo Manzo and Maria~F Garcia-Parajo.
\newblock A review of progress in single particle tracking: from methods to biophysical insights.
\newblock {\em Reports on Progress in Physics}, 78:124601, 2015.

\bibitem{cherstvy_anomalous_2021}
Andrey~G Cherstvy, Hadiseh Safdari, and Ralf Metzler.
\newblock Anomalous diffusion, nonergodicity, and ageing for exponentially and logarithmically time-dependent diffusivity: striking differences for massive versus massless particles.
\newblock {\em Journal of Physics D: Applied Physics}, 54:195401, 2021.

\bibitem{beaulieu_basis_2002}
Christian Beaulieu.
\newblock The basis of anisotropic water diffusion in the nervous system – a technical review.
\newblock {\em NMR in Biomedicine}, 15:435--455, 2002.

\bibitem{bechinger_active_2016}
Clemens Bechinger, Roberto Di~Leonardo, Hartmut Löwen, Charles Reichhardt, Giorgio Volpe, and Giovanni Volpe.
\newblock Active {Particles} in {Complex} and {Crowded} {Environments}.
\newblock {\em Reviews of Modern Physics}, 88:045006, 2016.

\bibitem{singh_sperm_2020}
Ajay~Vikram Singh, Mohammad Hasan~Dad Ansari, Mihir Mahajan, Shubhangi Srivastava, Shubham Kashyap, Prajjwal Dwivedi, Vaibhav Pandit, and Uma Katha.
\newblock Sperm {Cell} {Driven} {Microrobots}—{Emerging} {Opportunities} and {Challenges} for {Biologically} {Inspired} {Robotic} {Design}.
\newblock {\em Micromachines}, 11:448, 2020.

\bibitem{sokolov_swimming_2010}
Andrey Sokolov, Mario~M. Apodaca, Bartosz~A. Grzybowski, and Igor~S. Aranson.
\newblock Swimming bacteria power microscopic gears.
\newblock {\em Proceedings of the National Academy of Sciences}, 107:969--974, 2010.

\bibitem{vizsnyiczai_light_2017}
Gaszton Vizsnyiczai, Giacomo Frangipane, Claudio Maggi, Filippo Saglimbeni, Silvio Bianchi, and Roberto Di~Leonardo.
\newblock Light controlled {3D} micromotors powered by bacteria.
\newblock {\em Nature Communications}, 8:15974, 2017.

\bibitem{di_leonardo_bacterial_2010}
R.~Di~Leonardo, L.~Angelani, D.~Dell’Arciprete, G.~Ruocco, V.~Iebba, S.~Schippa, M.~P. Conte, F.~Mecarini, F.~De~Angelis, and E.~Di~Fabrizio.
\newblock Bacterial ratchet motors.
\newblock {\em Proceedings of the National Academy of Sciences}, 107:9541--9545, 2010.

\bibitem{patra_intelligent_2013}
Debabrata Patra, Samudra Sengupta, Wentao Duan, Hua Zhang, Ryan Pavlick, and Ayusman Sen.
\newblock Intelligent, self-powered, drug delivery systems.
\newblock {\em Nanoscale}, 5:1273--1283, 2013.

\bibitem{nelson_microrobots_2010}
Bradley~J. Nelson, Ioannis~K. Kaliakatsos, and Jake~J. Abbott.
\newblock Microrobots for {Minimally} {Invasive} {Medicine}.
\newblock {\em Annual Review of Biomedical Engineering}, 12:55--85, 2010.

\bibitem{peterkovic_optimising_2024}
Zane Peterkovic, Avinash Upadhya, Christopher Perrella, Admir Bajraktarevic, Ramses~Bautista Gonzalez, Megan Lim, Kylie~R. Dunning, and Kishan Dholakia.
\newblock Optimising image capture for low-light widefield quantitative fluorescence microscopy, 2024.

\bibitem{golibrzuch_application_2022}
Kai Golibrzuch, Sven Schwabe, Tianli Zhong, Kim Papendorf, and Alec~M. Wodtke.
\newblock Application of an {Event}-{Based} {Camera} for {Real}-{Time} {Velocity} {Resolved} {Kinetics}.
\newblock {\em The Journal of Physical Chemistry A}, 126:2142--2148, 2022.

\bibitem{cabriel_event-based_2023}
Clément Cabriel, Tual Monfort, Christian~G. Specht, and Ignacio Izeddin.
\newblock Event-based vision sensor for fast and dense single-molecule localization microscopy.
\newblock {\em Nature Photonics}, 17:1105--1113, 2023.

\bibitem{guo_eventlfm_2024}
Ruipeng Guo, Qianwan Yang, Andrew~S. Chang, Guorong Hu, Joseph Greene, Christopher~V. Gabel, Sixian You, and Lei Tian.
\newblock {EventLFM}: event camera integrated {Fourier} light field microscopy for ultrafast {3D} imaging.
\newblock {\em Light: Science \& Applications}, 13:144, 2024.

\bibitem{ren_event-based_2022}
Yugang Ren, Enrique Benedetto, Harry Borrill, Yelizaveta Savchuk, Molly Message, Katie O'Flynn, Muddassar Rashid, and James Millen.
\newblock Event-based imaging of levitated microparticles.
\newblock {\em Applied Physics Letters}, 121:113506, 2022.

\bibitem{ren_neuromorphic_2024}
Yugang Ren, Benjamin Siegel, Ronghao Yin, Muddassar Rashid, and James Millen.
\newblock Neuromorphic detection and cooling of microparticle arrays, 2024.
\newblock arXiv:2408.00661.

\bibitem{takatsuka_millisecond-scale_2024}
Susumu Takatsuka, Norio Miyamoto, Hidehito Sato, Yoshiaki Morino, Yoshihisa Kurita, Akinori Yabuki, Chong Chen, and Shinsuke Kawagucci.
\newblock Millisecond-scale behaviours of plankton quantified in vitro and in situ using the {Event}-based {Vision} {Sensor}.
\newblock {\em Ecology and Evolution}, 14:e70150, 2024.

\bibitem{gehrig_low-latency_2024}
Daniel Gehrig and Davide Scaramuzza.
\newblock Low-latency automotive vision with event cameras.
\newblock {\em Nature}, 629:1034--1040, 2024.

\bibitem{sadak_human_2024}
Ferhat Sadak, Edison Gerena, Charlotte Dupont, Rachel Lévy, and Sinan Haliyo.
\newblock Human {Sperm} {Detection} and {Tracking} using {Event}-based {Cameras} and {Unsupervised} {Learning}.
\newblock In {\em 2024 {International} {Conference} on {Manipulation}, {Automation} and {Robotics} at {Small} {Scales} ({MARSS})}, pages 1--6, Delft, Netherlands, 2024. IEEE.

\bibitem{gallego_event-based_2022}
Guillermo Gallego, Tobi Delbruck, Garrick Orchard, Chiara Bartolozzi, Brian Taba, Andrea Censi, Stefan Leutenegger, Andrew~J. Davison, Jorg Conradt, Kostas Daniilidis, and Davide Scaramuzza.
\newblock Event-{Based} {Vision}: {A} {Survey}.
\newblock {\em IEEE Transactions on Pattern Analysis and Machine Intelligence}, 44:154--180, 2022.

\bibitem{valli_super-resolution_2021}
Jessica Valli and Jeremy Sanderson.
\newblock Super-{Resolution} {Fluorescence} {Microscopy} {Methods} for {Assessing} {Mouse} {Biology}.
\newblock {\em Current Protocols}, 1:e224, 2021.

\bibitem{ershov_trackmate_2022}
Dmitry Ershov, Minh-Son Phan, Joanna~W. Pylvänäinen, Stéphane~U. Rigaud, Laure Le~Blanc, Arthur Charles-Orszag, James R.~W. Conway, Romain~F. Laine, Nathan~H. Roy, Daria Bonazzi, Guillaume Duménil, Guillaume Jacquemet, and Jean-Yves Tinevez.
\newblock {TrackMate} 7: integrating state-of-the-art segmentation algorithms into tracking pipelines.
\newblock {\em Nature Methods}, 19:829--832, 2022.

\bibitem{tinevez_trackmate_2017}
Jean-Yves Tinevez, Nick Perry, Johannes Schindelin, Genevieve~M. Hoopes, Gregory~D. Reynolds, Emmanuel Laplantine, Sebastian~Y. Bednarek, Spencer~L. Shorte, and Kevin~W. Eliceiri.
\newblock {TrackMate}: {An} open and extensible platform for single-particle tracking.
\newblock {\em Methods}, 115:80--90, 2017.

\bibitem{dhont_introduction_1996}
Jan K.~G. Dhont.
\newblock {\em An introduction to dynamics of colloids}.
\newblock Elsevier, Amsterdam, Netherlands New York, 1996.

\bibitem{qian_single_1991}
H.~Qian, M.P. Sheetz, and E.L. Elson.
\newblock Single particle tracking. {Analysis} of diffusion and flow in two-dimensional systems.
\newblock {\em Biophysical Journal}, 60:910--921, 1991.

\bibitem{michalet_mean_2010}
Xavier Michalet.
\newblock Mean square displacement analysis of single-particle trajectories with localization error: {Brownian} motion in an isotropic medium.
\newblock {\em Physical Review E}, 82:041914, 2010.

\bibitem{michalet_optimal_2012}
Xavier Michalet and Andrew~J. Berglund.
\newblock Optimal diffusion coefficient estimation in single-particle tracking.
\newblock {\em Physical Review E}, 85:061916, 2012.

\bibitem{ernst_measuring_2013}
Dominique Ernst and Jürgen Köhler.
\newblock Measuring a diffusion coefficient by single-particle tracking: statistical analysis of experimental mean squared displacement curves.
\newblock {\em Phys. Chem. Chem. Phys.}, 15:845--849, 2013.

\bibitem{carbajal-tinoco_asymmetry_2007}
Mauricio~D. Carbajal-Tinoco, Ricardo Lopez-Fernandez, and José~Luis Arauz-Lara.
\newblock Asymmetry in {Colloidal} {Diffusion} near a {Rigid} {Wall}.
\newblock {\em Physical Review Letters}, 99:138303, 2007.

\bibitem{boitrelle_sixth_2021}
Florence Boitrelle, Rupin Shah, Ramadan Saleh, Ralf Henkel, Hussein Kandil, Eric Chung, Paraskevi Vogiatzi, Armand Zini, Mohamed Arafa, and Ashok Agarwal.
\newblock The {Sixth} {Edition} of the {WHO} {Manual} for {Human} {Semen} {Analysis}: {A} {Critical} {Review} and {SWOT} {Analysis}.
\newblock {\em Life}, 11:1368, 2021.

\bibitem{world_health_organisation_laboratory_2021}
W.H.O.
\newblock {\em Laboratory {Manual} for the {Examination} and {Processing} of {Human} {Semen}}.
\newblock World Health Organisation, 2021.

\bibitem{van_der_horst_computer_2020}
Gerhard Van Der~Horst.
\newblock Computer {Aided} {Sperm} {Analysis} ({CASA}) in domestic animals: {Current} status, three {D} tracking and flagellar analysis.
\newblock {\em Animal Reproduction Science}, 220:106350, 2020.

\bibitem{vareasanchezGeometricMorphometricsRodent2013}
María Varea~Sánchez, Markus Bastir, and Eduardo R.~S. Roldan.
\newblock Geometric {Morphometrics} of {Rodent} {Sperm} {Head} {Shape}.
\newblock {\em PLoS ONE}, 8:e80607, 2013.

\bibitem{wilson-leedy_development_2007}
Jonas~G. Wilson-Leedy and Rolf~L. Ingermann.
\newblock Development of a novel {CASA} system based on open source software for characterization of zebrafish sperm motility parameters.
\newblock {\em Theriogenology}, 67:661--672, 2007.

\bibitem{amann_computer-assisted_2014}
Rupert~P. Amann and Dagmar Waberski.
\newblock Computer-assisted sperm analysis ({CASA}): {Capabilities} and potential developments.
\newblock {\em Theriogenology}, 81:5--17.e3, 2014.

\bibitem{alquezar-baeta_opencasa_2019}
Carlos Alquézar-Baeta, Silvia Gimeno-Martos, Sara Miguel-Jiménez, Pilar Santolaria, Jesús Yániz, Inmaculada Palacín, Adriana Casao, José~Alvaro Cebrián-Pérez, Teresa Muiño-Blanco, and Rosaura Pérez-Pé.
\newblock {OpenCASA}: {A} new open-source and scalable tool for sperm quality analysis.
\newblock {\em PLOS Computational Biology}, 15:e1006691, 2019.

\bibitem{goodson_classification_2011}
Summer~G. Goodson, Zhaojun Zhang, James~K. Tsuruta, Wei Wang, and Deborah~A. O'Brien.
\newblock Classification of {Mouse} {Sperm} {Motility} {Patterns} {Using} an {Automated} {Multiclass} {Support} {Vector} {Machines} {Model}.
\newblock {\em Biology of Reproduction}, 84:1207--1215, 2011.

\bibitem{yazdan_parast_viscous_2024}
Farin Yazdan~Parast, Shibani Veeraragavan, Avinash~S. Gaikwad, Sushant Powar, Ranganathan Prabhakar, Moira~K. O'Bryan, and Reza Nosrati.
\newblock Viscous {Loading} {Regulates} the {Flagellar} {Energetics} of {Human} and {Bull} {Sperm}.
\newblock {\em Small Methods}, 8:2300928, 2024.

\bibitem{yazdan_parast_cooperative_2023}
Farin Yazdan~Parast, Avinash~S. Gaikwad, Ranganathan Prabhakar, Moira~K. O’Bryan, and Reza Nosrati.
\newblock The cooperative impact of flow and viscosity on sperm flagellar energetics in biomimetic environments.
\newblock {\em Cell Reports Physical Science}, 4:101646, 2023.

\bibitem{nosrati_two-dimensional_2015}
Reza Nosrati, Amine Driouchi, Christopher~M. Yip, and David Sinton.
\newblock Two-dimensional slither swimming of sperm within a micrometre of a surface.
\newblock {\em Nature Communications}, 6:8703, 2015.

\bibitem{willert_event-based_2023}
Christian~E. Willert.
\newblock Event-based imaging velocimetry using pulsed illumination.
\newblock {\em Experiments in Fluids}, 64:98, 2023.

\bibitem{cao_overview_2020}
Keyan Cao, Yefan Liu, Gongjie Meng, and Qimeng Sun.
\newblock An overview on edge computing research.
\newblock {\em IEEE Access}, 8:85714--85728, 2020.

\bibitem{lin_all-optical_2018}
Xing Lin, Yair Rivenson, Nezih~T. Yardimci, Muhammed Veli, Yi~Luo, Mona Jarrahi, and Aydogan Ozcan.
\newblock All-optical machine learning using diffractive deep neural networks.
\newblock {\em Science}, 361:1004--1008, 2018.

\bibitem{yan_fourier-space_2019}
Tao Yan, Jiamin Wu, Tiankuang Zhou, Hao Xie, Feng Xu, Jingtao Fan, Lu~Fang, Xing Lin, and Qionghai Dai.
\newblock Fourier-space {Diffractive} {Deep} {Neural} {Network}.
\newblock {\em Physical Review Letters}, 123:023901, 2019.

\bibitem{colburn_optical_2019}
Shane Colburn, Yi~Chu, Eli Shilzerman, and Arka Majumdar.
\newblock Optical frontend for a convolutional neural network.
\newblock {\em Applied Optics}, 58:3179--3186, 2019.

\bibitem{zheng_meta-optic_2022}
Hanyu Zheng, Quan Liu, You Zhou, Ivan~I. Kravchenko, Yuankai Huo, and Jason Valentine.
\newblock Meta-optic accelerators for object classifiers.
\newblock {\em Science Advances}, 8:eabo6410, 2022.

\bibitem{qiu_spiral_2018}
Xiaodong Qiu, Fangshu Li, Wuhong Zhang, Zhihan Zhu, and Lixiang Chen.
\newblock Spiral phase contrast imaging in nonlinear optics: seeing phase objects using invisible illumination.
\newblock {\em Optica}, 5:208, 2018.

\bibitem{swartz_broadband_2024}
Brandon~T. Swartz, Hanyu Zheng, Gregory~T. Forcherio, and Jason Valentine.
\newblock Broadband and large-aperture metasurface edge encoders for incoherent infrared radiation.
\newblock {\em Science Advances}, 10:eadk0024, 2024.

\end{thebibliography}

% Use the following code if you wish to generate your bibliography with BibTeX;
% replace the string "MSP-template" below with the name(s) of
% the BibTeX data base(s) you want to use.
% The resulting bibliography-output (the content of the .bbl file)
% must be pasted back into this file before submission.
% Please also include your BibTeX data base file(s) in your submission
% so that we can re-run BibTeX if necessary.

\end{document}